\documentclass[aps,pra,twocolumn,groupedaddress]{revtex4}
\usepackage{graphicx}
\usepackage{amsmath}
\usepackage{amsfonts}
\usepackage{color}
\usepackage{bm}
\newcommand{\ket}[1]{\ensuremath{\left|{#1}\right\rangle}}
\newcommand{\bra}[1]{\ensuremath{\left\langle{#1}\right |}}

\newcommand{\oper}[1]{\mathbf{\mathsf{#1}}}

\newcommand{\brm}[1]{\ensuremath{\mathbf{#1}}}

\newcommand{\sinc}{\ensuremath{{\mathrm{sinc}}}}

\newcommand{\mmod}{\ensuremath{{\mathrm{mod}}}}

\newcommand{\rh}{\boldsymbol{\rho}}
\newcommand{\rhb}{\boldsymbol{ \bar{\rho}}}
\newcommand{\bq}{{\bf q}}
\newcommand{\bqb}{\bar{\bf q}}
\newcommand{\beq}{\begin{equation}}
\newcommand{\eeq}{  \end{equation}}
\newcommand{\bea}{\begin{eqnarray}}
\newcommand{\eea}{  \end{eqnarray}}
\newcommand{\bit}{\begin{itemize}}
\newcommand{\eit}{  \end{itemize}}

\begin{document}


\title{Propagation of transverse intensity correlations of a two-photon state}

\author{D. S. Tasca}
\email[]{tasca@if.ufrj.br} \affiliation{Instituto de F\'{\i}sica,
Universidade Federal do Rio de Janeiro, Caixa Postal 68528, Rio de
Janeiro, RJ 21941-972, Brazil}
\author{S. P. Walborn}
\affiliation{Instituto de F\'{\i}sica, Universidade Federal do Rio
de Janeiro, Caixa Postal 68528, Rio de Janeiro, RJ 21941-972,
Brazil}
\author{F. Toscano}
\affiliation{ Funda\c{c}\~ao Centro de Ci\^encias e
              Educa\c{c}\~ao Superior a Dist\^ancia do Estado
              do Rio de Janeiro,
              20943-001 Rio de Janeiro, Brazil}
 \affiliation{Instituto de F\'{\i}sica, Universidade Federal do Rio
              de Janeiro, Caixa Postal 68528, Rio de Janeiro, RJ 21941-972,
              Brazil}

\author{P. Pellat-Finet}
\affiliation{Groupe d'Optique Th\'eorique et Appliqu\'ee, LMAM, Universit\'e de Bretagne Sud, B. P. 92116, Lorient cedex, France}
\affiliation{D\'epartement d'Optique, UMR CNRS 6082, \'Ecole Nationale Sup\'erieure des T\'el\'ecommunications de Bretagne, France}


\author{P. H. Souto Ribeiro}
\email[]{phsr@if.ufrj.br} \affiliation{Instituto de F\'{\i}sica,
Universidade Federal do Rio de Janeiro, Caixa Postal 68528, Rio de
Janeiro, RJ 21941-972, Brazil}
\date{\today}

\begin{abstract}
The propagation of transverse spatial correlations of photon pairs through arbitrary first-order linear  optical systems is studied experimentally and theoretically using the fractional Fourier transform.  Highly-correlated photon pairs in an EPR-like state are produced by spontaneous parametric down-conversion and subject to optical fractional Fourier transform systems.  It is shown 
 that the joint detection probability can display either correlation, anti-correlation, 
or no correlation, depending on the sum of the orders $\alpha$ and $\beta$ of the transforms of the down-converted photons. We present analytical results for the propagation of the perfectly correlated EPR state, and numerical results for the propagation of the two-photon state produced from parametric down-conversion.  We find good agreement between theory and experiment.       
  \end{abstract}


\maketitle


\section{Introduction}
The discussion about non-local correlations between properties of two separated
particles began in part with the famous EPR paper \cite{epr35}, in which Einstein, Podolsky and Rosen showed that the position
and momentum of two correlated particles could be used to construct a paradox between
quantum theory and intuitive concepts like locality and the reality of
physical properties.  Continuous variable (CV) entangled states similar to the EPR state appear in a number of physical systems, including field-quadrature correlations of two modes of the electromagnetic field \cite{ou92,silberhorn01,villar05,takeno07}, spatial variables of pairs of photons \cite{howell04,dangelo04}, and others \cite{julsgaard01,bowen02,bowen03,fedorov05}.    This has allowed for experimental realization of the original gedanken experiment proposed by EPR \cite{ou92,howell04,dangelo04}.  CV entanglement of the EPR type has been shown to be useful for a number of quantum information tasks \cite{braunstein05}.  One benefit to the study and use of CV's is access to a Hilbert space of larger dimension, which is advantageous for quantum cryptography \cite{bechmann00a,bourennane01} and fundamental tests of quantum mechanics \cite{collins02}.      
\par
EPR-like spatial correlations can be identified by 
the violation of the inequality \cite{reid88,reid08}
\begin{equation}\label{EQ:EPRinequality}
\Delta^2(\rho_{1}{|\rho_{2}})\Delta^2
(q_{1}|q_{2})>\frac{1}{4},
\end{equation}
 where $\Delta^2(\rho_{1}|\rho_{2})$ 
 represents the uncertainty in variable $\rho_1$ of system 1 conditioned upon measurement of system 2 
at $\rho_2$. $\Delta^2(\rho_{1}|\rho_{2})$ is the variance of the conditional probability $P(\rho_{1}|\rho_{2})$ for a fixed value of $\rho_2$.
Similarly,  $\Delta^2(q_{1}|q_{2})$ 
is the variance
of the conditional probability $P(q_{1}|q_{2})$, where $q_1$ and $q_2$ are the Fourier conjugate variables of $\rho_1$ and $\rho_2$. 
If inequality (\ref{EQ:EPRinequality}) is violated, one could infer
either $\rho_{1}$ or $q_{1}$ from conditional measurements 
$\rho_{2}$ or $q_{2}$, with less uncertainty than the
Heisenberg uncertainty principle would allow.  In recent experiments \cite{howell04,dangelo04}, measurements of this
type were performed in the coincidence counting regime using photons from spontaneous parametric down-conversion (SPDC).  The 
transverse position and momentum were determined by measuring the intensity distributions in the near and far
field, respectively.  As inequality \eqref{EQ:EPRinequality} deals with EPR non-locality, it is generally more restrictive than those involving variances of center of mass and relative variables, which identify non-separability of continuous variable systems \cite{duan00,mancini02}.      
\par
The spatial correlations of photon pairs produced by  SPDC present a rich playground to investigate CV correlations with relatively simple linear optical systems \cite{strekalov94,ribeiro94b,pittman96,monken98a,abouraddy01,torres03a,yarnall07a}.  In SPDC sources, photon pairs generally display an intensity correlation in the near field (source), due to the localized emission of the photon pair:  the photons are ``born" from the same pump photon, so both photons are detected at nearly the same position in the source plane.  As the entangled two-photon state propagates, this spatial correlation evolves to an anti-correlation in the far field.  Consequently, if photon 1 is detected at position $\rh$ in the far-field, photon 2 will be found near $-\rh$.  The far-field anti-correlations are due to the phase matching (momentum conservation) in the non-linear SPDC interaction.  The spatial correlation in the near-field and anti-correlation in the far-field have been previously observed in Ref. \cite{almeida05}. The switch from a near-field correlation to a far-field anti-correlation raises the question as to what type of correlation is present at intermediate distances in between the near and far-field regions.  Recently, Chan et al. \cite{chan07} showed that the correlations can ``migrate" entirely to the phase of the two-photon wave function, and consequently the conditional intensity distribution may display no correlation at all.   In Ref. \cite{tasca2008}, it was shown that it is always possible to detect transverse entanglement performing 
only intensity correlation measurements, when an arbitrary propagation is applied to each of the entangled photons.    
  \par
The propagation of the transverse spatial structure of an optical field can be accurately described by the Fractional Fourier Transform (FRFT) \cite{ozaktas01}.  This is true for any first-order linear optical system. That includes free-space propagation alone \cite{alieva94,pellat-finet94a,pellat-finet94b}, and also optical systems consisting of lenses and free space \cite{lohmann93,ozaktas01}, provided that one chooses the appropriate scaling of the transverse coordinates.  The FRFT is parameterized by an angle $\alpha$, so that $\alpha=0$ corresponds to an identity operation and $\alpha=\pi/2$ is the usual Fourier transform.  With proper scaling of the coordinates, the FRFT is additive, so that consecutive FRFT's $\mathcal{F}_{\alpha}$ and $\mathcal{F}_{\beta}$ can be written as $\mathcal{F}_{\alpha+\beta}$.  This allows one to associate an overall FRFT to an arbitrary first-order linear optical system.  
\par
In the present work, we study the transverse EPR correlations of propagating SPDC photon pairs using the FRFT. We show theoretically and experimentally that the presence of
EPR intensity correlation, anti-correlation or no correlation depends on the sum of the orders $\alpha$ and $\beta$ of the applied FRFT transforms in each of the down converted photons.
In this way, it is possible to engineer the spatial intensity correlations through the application of optical FRFTs to the entangled down-converted photons.
The FRFT describes a canonical rotation in phase space, and applies to any pair of conjugate variables, such as time-frequency \cite{ozaktas01} or field quadratures \cite{braunstein05}.
Thus, the conclusions drawn here are also relevant to other physical systems.
\par  
In section \ref{sec:hilb},  we review the connection between the Hilbert space associated to the spatial variables of a single and two-photon field
and the Hilbert spaces of point particles with two degrees of freedom. This allows us to apply the
usual quantum formalism for point particles in the description of the spatial properties of single and two-photon states.
In section \ref{sec:FRFTprop}, we discuss the propagation of photons  through  first-order linear optical systems and the use of the FRFT in this description.  Section \ref{sec:twophoton} introduces the type of two photon state
typical of the SPDC process, and discusses the propagation of transverse correlations under FRFT operations.  In section \ref{sec:exp} we present an experiment and results which are well described by the theoretical results presented in 
section \ref{sec:twophoton}. 
Finally,  we provide some concluding remarks in section \ref{sec:conclusion}.

\section{Single and two-photon states}
\label{sec:hilb}
Here we focus on the spatial structure of a single or two-photon field.  Thus, for simplicity, we will assume that the fields are paraxial, monochromatic and have well defined polarization.  The Hilbert space ${\cal H}_{1}$ describing the transverse spatial degrees of freedom of a single-photon state $|\psi\rangle$ is spanned
by the basis $\{\ket{\rhb}\equiv \hat{a}^{\dagger}(\rhb)|0\rangle\}$, where
$|0\rangle$ is the vacuum state.  An arbitrary pure state is then
\begin{equation}
\ket{\psi}=\int d\rhb \,w(\rhb )\ket{\rhb}, 
\end{equation}
where
$\rhb \equiv (\bar{\rho}_x,\bar{\rho}_y)$ is the transverse position
and $w(\rhb)$
is the transverse wavefunction or detection amplitude.
The basis states $\{\ket{\rhb}\}$ correspond in second quantization to unnormalized states of one photon
at position $\rhb$.
It is possible to establish an isomorphism between ${\cal H}_{1}$ 
and the Hilbert space  spanned by position eigenstates of a
two-dimensional position operator $\hat{\rhb}\equiv(\hat{\bar{\rho}}_x,\hat{\bar{\rho}}_y)$
if one specifies the action of this operator on the basis states as:  
$\hat{\rhb}|\rhb \rangle=\rhb |\rhb\rangle$.
\par
Alternatively, ${\cal H}_1$ is spanned
by the basis $\{\ket{\bqb}\equiv \hat{a}^{\dagger}(\bqb)|0\rangle\}$, where
\begin{equation}
\label{FTadagger}
\hat{a}^{\dagger}(\bqb)=\int d\rhb\, e^{i\rhb\cdot\bqb}\,\hat{a}^{\dagger}(\rhb)
\end{equation}
and $\bqb\equiv(\bar{q}_x,\bar{q}_y)$ are the transverse components
of the wave vector ${\brm{k}}$.
In this basis the wavefunction $v(\bqb)$
is the angular spectrum of the photon
field, and is obtained by a Fourier transform of the detection
amplitude $w(\rhb)$.
Again, it is posible to establish an isomorphism between ${\cal H}_1$
and the space spanned by momentum eigenstates of a
two-dimensional momentum operator $\hat{\bqb}$$\equiv(\hat{\bar{q}}_x,\hat{\bar{q}}_y)$
if the action of this operator on the basis
states is:  $\hat{\bqb}|\bqb \rangle=\bqb |\bqb\rangle$.
Because the two bases $\{\ket{\rhb}\}$ and $\{\ket{\bqb}\}$ are related via a Fourier transform similar to   the one in Eq. (\ref{FTadagger}),
the position and momentum operators satisfy the canonical
commutation relations $[\hat{\bar{\rho}}_k,\hat{\bar{q}}_l]=i\delta_{k,l}\hat{\mathbb{I}}$, where $k,l=\{x,y\}$.
Thus, at the level of quantum kinematics, there is an isomorphism between 
the Hilbert space corresponding to transverse spatial degrees of freedom of single-photon states
and the Hilbert space of  quantum states of a point particle with two degrees of freedom.
The equivalence between the classical paraxial wave optics and the nonrelativistic
quantum mechanics of two-dimensional point particles is well known  
\cite{marcuse82}, and also allows one to establish the isomorphism at the level of quantum dynamics.  
In fact, for paraxial propagation of the photons along an optical axis $z$,  
the wave equation that governs the evolution of the wavefunction $w(\rhb )\equiv \langle \rhb|w\rangle$, is 
a time dependent Schr\"odinger equation where the length variable $z$ plays the role
of time and the wavelength, $\lambda$, of the photons plays the role of Planck's constant  
\footnote{ Note that in the paraxial approximation the transverse components of the wave vector $\vec{k}$ are
$q_l\approx k\theta_l$ ($l=x,y$ and $k\equiv |\vec{k}|=2\pi/\lambda$) where $\theta_l\ll 2\pi$ are the angles between $\vec{k}$
and the $z$ axis. Thus, considering $\lambda/2\pi$ as the analogous of  $\hbar$  
we can write $[\hat{\rho}_k,\hat{\theta}_l]=i (\lambda/2\pi) \delta_{k,l}$
or $[\hat{\rho}_k,\hat{q}_l]=i\delta_{k,l}$.}.  The analogy between paraxial wave propagation and non-relativistic quantum mechanics of a point particle has been well explored \cite{gloge69,bacry81,stoler81,marcuse82,ozaktas01,dragoman02}.

The Hilbert space describing the transverse spatial degrees of freedom of  two-photon states is simply
the tensor product ${\cal H}_{1}\otimes {\cal H}_{2} $ between the Hilbert spaces of one-photon states.
Thus, ${\cal H}_{1}\otimes {\cal H}_{2}$ is isomorphic to the Hilbert space associated to two distinguishable point particles, each one with two degrees
of freedom.  We assume that the photons are distinguishable since in principle they could be distinguished by their longitudinal direction of propagation or their polarization. Therefore,  an arbitrary  two-photon pure state can be written as
\begin{equation}
\ket{\Psi} = \iint d\rhb_1 d\rhb_2 \Psi(\rhb_1,\rhb_2) \ket{\rhb_1}_1\ket{\rhb_2}_2, 
\label{eq:state}
\end{equation}
where $\Psi(\rhb_1,\rhb_2)= \bra{\rhb_1,\rhb_2}\Psi\rangle$ is the normalized wave function and
$ \ket{\rhb_1}_1$ and $\ket{\rhb_2}_2$ are position eigenstates for photons $1$ and $2$, respectively.  Here it is assumed that the paraxial approximation has been applied along two distinct $z$ axes, one for each single-photon field.

\subsection{Propagation as a Fractional Fourier Transform}
\label{sec:FRFTprop}

The most common optical systems are  
first-order linear systems (also called quadratic-phase systems),  which  are composed essentially  
of sections of free space and thin lenses centered on the propagation ($z$) axis \cite{ozaktas01}.
Paraxial propagation  in these systems
is particularly simple: the paraxial wave equation corresponds to a Schr\"odinger equation
associated with a quadratic hamiltonian, so the evolution of the phase space operators 
are simply given by $(\hat{\rhb},\hat{\bqb})^T=\mbox{$\hat{U}^{\dagger}$} (\hat{\rhb},\hat{\bqb})^T \mbox{$\hat{U}$}=\mbox{$M$}(\hat{\rhb},\hat{\bqb})^T$ (where $\hat{U}$ is
the evolution operator associated with the quadratic hamiltonian, and T means transposition).
The symplectic matrix $M$ is the ray matrix that stems from geometrical optics applied to the system.
For example, in the case of only free propagation the evolution is associated with the hamiltonian of a free particle
and the matrix $M$ represent a linear canonical transformation that correspond to a shear in the direction of
the  transverse momentum \cite{ozaktas01}.

\begin{figure}
\centering
\includegraphics*[width=7cm]{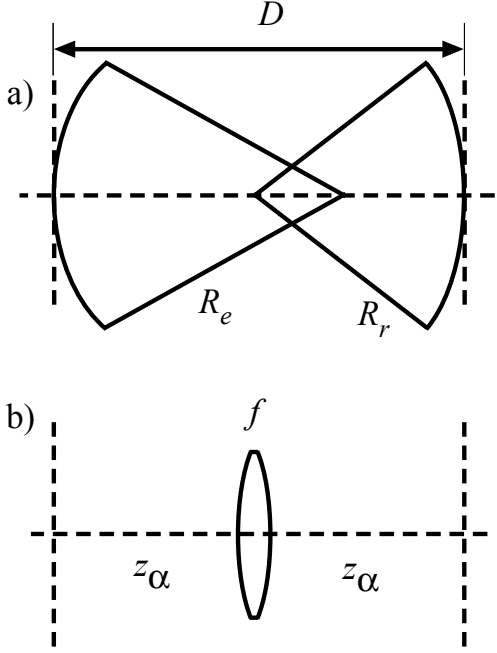}
\caption{\label{FIG:FRFTs} a) In free diffraction, the field on the curved surface with radius $R_r$ can be described as the FRFT of the field on surface of radius $R_e$ with properly scaled coordinates.  b)  The FRFT can be implemented with a simple lens symmetrical system \cite{lohmann93}.}
\end{figure}

A great simplification and systematization in the description of evolution through first-order optical systems 
is gained by using dimensionless variables $\rh=\rhb/s$ and $\bqb=s\bq$, where the real number
$s$ has the dimension of a length, and is generally 
a function of the properties of the physical system.
In this case, free space propagation can be described in the paraxial approximation with the help of the FRFT \cite{pellat-finet94a,pellat-finet94b,ozaktas01}.
This is due to the fact that the paraxial Fresnel diffraction integral, which 
relates the light signal between two transverse planes in free space, can be expressed
using a FRFT if we use dimensionless coordinates.  The more general case occurs when we choose different parameters $s$
 at the
input and the output planes.   However, in order to identify the tranverse position and momentum coordinates 
at these planes as belonging to the same phase space, one must use the same parameter $s$.
FIG. \ref{FIG:FRFTs} a) illustrates identification of the FRFT with propagation through free space.  The diffraction of light from a spherical cap emitter with radius of curvature 
$R_e=-R<0$ to a spherical cap receiver with radius of curvature $R_r=R>0$ at a distance $z$
from the emitter can be expressed as \cite{pellat-finet94a,pellat-finet94b}
\beq
\label{FRFTinput-ouput}
\phi_r(\rh)=e^{-i\,\alpha/2}\mathcal{F}_{\alpha}[\phi_e(\rh)],
\eeq
where $\phi_e(\rh)=\exp(-ik|s\rho|^2/2R_e)\varphi_e(\rh)$ and $\phi_r(\rh)=\exp(-i k|s\rho|^2/2R_r)\varphi_r(\rh)$ ($k\equiv |\brm{k}|=2\pi/\lambda$). $\varphi_e(\rh)$ and $\varphi_r(\rh)$ are the wavefunctions at planes of observations tangent to the emitter's and receiver's
spherical caps at its vertex point. Here we call the angle $0<\alpha<\pi$  the order of the FRFT. This order
$\alpha\equiv\alpha(R,z)$
and the adimensionalization parameter
$s\equiv s(R,z)$ can be calculated from the relations $s=\sqrt{z/k}\;(1-g^2)^{-1/4}$ and $g\equiv1-z/R=\cos\alpha$.
Alternatively, given the parameter $s$ and the distance $z$ we can estimate the FRFT's order $\alpha\equiv\alpha(s,z)$
and the radius of curvature $R\equiv R(s,z)$.   
It is important to note that the quadratic phase factors that maps
the wavefunctions at the spherical caps to the wavefunctions at their tangent planes are not important if we are concerned only  
with intensity  measurements at these planes.

The description  of the propagation of photons through first-order optical systems with the help of the FRFT is completed
if we use Eq.(\ref{FRFTinput-ouput}) in the section of free propagation, and for the 
action of thin lens we multiply the wavefunction at the plane of the lens by the phase factor
$\exp({-ik|s\rho|^2/2f})$, where $f$ is the focal length of the lens. 
It is important to maintain the same dimensionless parameter $s$ along the entire optical system in order
to use the additivity property:  $\mathcal{F}_{\alpha+\beta}=\mathcal{F}_{\alpha}\circ \mathcal{F}_{\beta}$ of the FRFT. 
This is the mechanism behind the implementation of a FRFT between two planar surfaces
with the optical systems reported  in \cite{lohmann93}, where  
$\varphi_r(\rh)=\mathcal{F}_{\alpha}[\varphi_e(\rh)]$.
In the experiment reported in section \ref{sec:exp},  we perform FRFT's using  the ``type I" symmetric lens system
configuration, which is illustrated in FIG. \ref{FIG:FRFTs} b). This FRFT system was originally reported in \cite{lohmann93} and is also discussed in detail in \cite{ozaktas01}.  This FRFT system consists of a
lens of focal length $f$ placed symmetrically
between the input and output planes, at a distance $z_{\alpha}$ from each. One can apply either Fourier optics or geometric optics to verify that this system corresponds to a FRFT.  Specifically, it is necessary to define the fractional focal length $f^{\prime}=f
\sin\alpha$,  and impose that the focal length $f$ and the distance of
propagation $z_{\alpha}$ before and after the lens are related to the order $\alpha$ of the FRFT via the
relation $z_{\alpha}=2f\sin^2(\alpha/2)$.     
The dimensionless position and momentum
coordinates for this kind of system are
$\rh = \sqrt{{k}/{f^{\prime}}}{\rhb}$ and
$\bq = \sqrt{{f^{\prime}}/{k}}{\bqb}$, where  $\rhb$, $\bqb$ are the dimensional variables.
In operator formalism, the evolution with a FRFT is associated with the fractional Fourier operator defined as \cite{ozaktas01}
\begin{equation}
\label{EQ:FRFToperator}
\hat{\mathcal{F}}_{\alpha} \equiv e^{i \alpha/2}
\exp{\left(-i\alpha\frac{\hat{\rh}^2+\hat{\bq}^2}{2}\right)}\;\;,
\end{equation}
where $\hat{\rh}$ and $\hat{\bq}$ are the dimensionless position and momentum operators.  
This operator is equivalent to the evolution operator of the quantum harmonic oscillator, with the hamiltonian defined as $\oper{H}={(\hat{\rh}^2+\hat{\bq}^2)}/{2}$.  
Under the FRFT operator of order $\alpha$, the single photon state $\ket{\varphi_{0}}$ evolves to $\ket{\varphi_{\alpha}}=\hat{\mathcal{F}}_{\alpha}\ket{\varphi_0}$.
The FRFT of the wave function $\varphi_{\alpha}(\rh)=\langle\rh|\varphi_{\alpha}\rangle$ is then given by \cite{pellat-finet94a}
\beq
\label{EQ:FRFTtransformation}
\varphi_{\alpha}(\rh) =  \int d\rh^{\prime}
 \bra{\rh}\hat{\mathcal{F}}_{\alpha}\ket{\rh^{\prime}}
\varphi_{0}(\rh^\prime),
\eeq
where the kernel is
\begin{eqnarray}
\bra{\rh}\hat{\mathcal{F}}_{\alpha}\ket{\rh^{\prime}}&\equiv&
A_{\alpha}
\exp\left(i\frac{\cot\alpha}{2}{\rho^{\prime}}^2\right) \nonumber \\
& \times &\exp\left(i\frac{\cot \alpha
}{2} {\rho}^2\right) \exp{\left(-i
\frac{\rh\cdot\rh^{\prime}}{\sin\alpha}\right)},
\label{eq:kernelFRFT}
\end{eqnarray}
for $0<|\alpha|<\pi$.  Here $A_{\alpha}=-i \exp(i\alpha/2)/(2\pi|\sin\alpha|)$.  Taking the limit $\alpha\rightarrow 0$ (or $\alpha\rightarrow 2 \pi$), one can show that $\bra{\rh}\hat{\mathcal{F}}_{\alpha}\ket{\rh^{\prime}}=\delta(\rh-\rh^{\prime})$ and similarly
$\bra{\rh}\hat{\mathcal{F}}_{\alpha}\ket{\rh^{\prime}}=\delta(\rh+\rh^{\prime})$ for
$\alpha \rightarrow \pm\pi$ \cite{ozaktas01}.
When $\alpha=\pi/2$, the FRFT reduces to the common Fourier transform.   When  $\alpha$ does not lie in the interval $0<|\alpha|<\pi$, Eq. \eqref{eq:kernelFRFT} accurately represents the FRFT kernel provided one replaces $\alpha$ with its value modulo $2\pi$. The transverse position and wave-vector operators
evolved under the action of FRFT are
\begin{equation}
\label{EQ:RhoqEvolved}
\left(  \begin{array}{c}
   \hat{\rh}_{\alpha}\\
   \hat{\bq}_{\alpha}
  \end{array}\right)=
  \hat{\mathcal{F}}^{\dagger}_{\alpha}
   \left(\begin{array}{c}
  \hat{\rh}\\
   \hat{\bq}
  \end{array}\right)
  \hat{\mathcal{F}}_{\alpha}
  =
  \left(\begin{array}{cc}
  \cos\alpha& \sin\alpha\\
  -\sin\alpha & \cos\alpha
  \end{array}\right)
  \left(\begin{array}{c}
  \hat{\rh}\\
   \hat{\bq}
  \end{array}\right)
\end{equation}
which illustrates the fact that $\hat{\mathcal{F}}_{\alpha}$ corresponds to rotation of angle $\alpha$ in phase space \cite{lohmann93,ozaktas01}.

\section{The entangled two photon state}
\label{sec:twophoton}

Let us consider now a pure two photon state whose wavefunction in dimensionless coordinates is of the form
\beq
\label{EQ:SpecialTwoPhotonState}
\Psi(\rh_1,\rh_2)=f(\rh_1+\rh_2) g(\rh_1-\rh_2)\;.
\eeq
This  state is generally correlated, provided that $f(\rh)$ and $g(\rh)$ are not identical Gaussian functions.  
Here it is assumed that $f$ and $g$ are normalized with respect to $\rh_1$, $\rh_2$.  The state \eqref{EQ:SpecialTwoPhotonState} can be readily produced in a number of physical processes \cite{fedorov05,torres1}.  It is representative of the two photon state at the face of the SPDC crystal, for example, provided that the pump and down-converted fields are polarized and nearly monochromatic \cite{torres2}.  In this case $f$ is given by the spatial profile of the pump field and $g$ is the Fourier transform of the phase matching function $G(\brm{q})=\sqrt{2L/K\pi^2}\sinc(L|\brm{q}|^2/4K)$
\cite{walborn03a}, where $K$ is the wave number of the pump beam.  
In many experimental situations, $G(\brm{q})$ and $g(\rh)$ can be approximated by Gaussian functions.     
In this case, assuming that the pump laser has a Gaussian profile, the position space wave function takes the form
\begin{equation}\label{EQ:TwopartWaveFuncPOS}
 \Psi(\rh_1,\rh_2)= \frac{1}{{\pi}\sigma_-\sigma_+}
    e^{-\frac{|\rh_1+\rh_2|^2}{4\sigma_+^2}}
   e^{-\frac{|\rh_1-\rh_2|^2}{4\sigma_-^2}}.
 \end{equation}
 Eq.
(\ref{EQ:TwopartWaveFuncPOS}) describes the field at the crystal face.
 In transverse wave-vector space the wave function is
\begin{equation}\label{EQ:TwopartWaveFuncMOM}
   \Psi(\brm{q}_1,\brm{q}_2)={\frac{\sigma_+ \sigma_-}{\pi}}
    e^{-\frac{\sigma_+^2}{4}|\bq_1+\bq_2|^2}
    e^{-\frac{\sigma_-^2}{4}|\bq_1-\bq_2|^2},
\end{equation}
which is obtained by taking the Fourier transform of the  wave function \eqref{EQ:TwopartWaveFuncPOS}.
Now let us suppose  that $\sigma_-\ll \sigma_+$, so that the photons exhibit a position correlation 
 and a momentum anticorrelation.  This is indeed what is generally produced in SPDC, in which it is not unusual to have $\sigma_- \sim \sigma_+/100$.  
 \subsection{Propagation of transverse correlations}
 \par
As discussed above, propagation of the down-converted fields can generally be described by a FRFT operation.  Let us assume that photon 1 propagates according to an $\alpha$-order FRFT along axis $z_1$, and photon 2 according to a $\beta$-order FRFT along axis $z_2$.  The state
$\ket{\Psi}$ after propagation is given by
\begin{equation}\label{EQ:EvolvedState}
\ket{\Psi_{\alpha,\beta}}=\hat{\mathcal{F}}^{(1)}_{\alpha} \otimes
\hat{\mathcal{F}}^{(2)}_{\beta}\ket{\Psi}.
\end{equation}
 The two-photon wave function then becomes  $\Psi_{\alpha,\beta}(\rh_1,\rh_2)=\langle \rh_1,\rh_2|\Psi_{\alpha,\beta}\rangle $, where  
 \begin{align}
\Psi_{\alpha,\beta}(\rh_1,\rh_2) = & \iint d\rh_1^{\prime}d\rh_2^{\prime}
 \bra{\rh_1}\hat{\mathcal{F}}_{\alpha}\ket{\rh_1^{\prime}} \times \nonumber \\
 & \bra{\rh_2}\hat{\mathcal{F}}_{\beta}\ket{\rh_2^{\prime}}
\Psi(\rh_1^\prime,\rh_2^\prime),
 \end{align}  
and the kernels are defined in Eq. \eqref{eq:kernelFRFT}.  To get a sense of the action of the FRFT's, let us consider the limiting case of an Einstein-Podolsky-Rosen (EPR) state, for which  $f(\rh)\sim \mathrm{constant}$ and $g(\rh)\sim \delta(\rh)$, giving 
$\Psi(\rh_1,\rh_2)=\delta(\rh_1-\rh_2)$.  This situation is approximated by the state produced by SPDC when the pump beam can be treated as a plane wave.  The EPR state is 
\begin{equation}
\ket{\Psi^{\mathrm{EPR}}} = \iint d\rh_1 d\rh_2 \delta(\rh_1-\rh_2) \ket{\rh_1}_1\ket{\rh_2}_2,   
\label{eq:EPR}
\end{equation} 
which presents a perfect correlation, since detection of photon $2$ at position $\rh$ projects photon $1$ onto a position eigenstate $\ket{\rh}$.  After FRFT's, the wave function $\Psi^{\mathrm{EPR}}_{\alpha,\beta}$ is 
\begin{align}
\Psi^{\mathrm{EPR}}_{\alpha,\beta}(\rh_1,\rh_2) = & A_{\alpha}A_{\beta} \exp\left(i\frac{\cot\alpha\rho_1^2+ \cot\beta\rho_2^2}{2}\right)\nonumber \\
& \int d\rh \exp\left(i\frac{\cot\alpha\rho^2}{2}\right)\exp\left(i\frac{\cot\beta\rho^2}{2}\right)  \nonumber
 \\ & \times
 \exp{\left [-i \rh \cdot \left ( \frac{\rh_1}{\sin\alpha}+\frac{\rh_2}{\sin\beta}\right) \right]}.
\end{align}
 Performing the integral, we have
 \begin{align}
\Psi^{\mathrm{EPR}}_{\alpha,\beta}(\rh_1,\rh_2) = & A_{\alpha+\beta} \exp\left[ i\frac{\cot(\alpha+\beta)}{2}\left(\rho_1^2 +\rho_2^2\right)\right ] \times \nonumber \\ 
& \exp \left[ -i \frac{\rh_1\cdot\rh_2}{\sin(\alpha+\beta)} \right],
 \end{align}
 which is the kernel of an FRFT of order $\alpha+\beta$ corresponding to propagation from an input plane (e.g. $\rh_1$) to an output plane (e.g. $\rh_2$).  
 The state $\ket{\Psi^{\mathrm{EPR}}_{\alpha,\beta}}$ is then  
  \begin{align}\label{eq:EPRevolved}
  \ket{\Psi_{\alpha,\beta}^{\mathrm{EPR}}}= A_{\alpha+\beta} &\iint d\rh_1 d\rh_2 \exp\left[ i\frac{\cot(\alpha+\beta)}{2}\left(\rho_1^2 +\rho_2^2\right) \right ] \times \nonumber \\ 
  & \exp \left[-i \frac{\rh_1\cdot\rh_2}{\sin(\alpha+\beta)} \right] \ket{\rh_1}_1\ket{\rh_2}_2.   
\end{align}
\par
Using the definition of the FRFT kernel \eqref{eq:kernelFRFT}, we note that whenever $\alpha+\beta = 0 \;(\mmod 2\pi)$, the original state \eqref{eq:EPR} is recovered.  That is, the EPR state \eqref{eq:EPR} is an eigenstate of operators of the type $\hat{\mathcal{F}_{\alpha}}\hat{\mathcal{F}}_{2\pi-\alpha}$, $\hat{\mathcal{F}_{\alpha}}\hat{\mathcal{F}}_{4\pi-\alpha}$, etc.   
When  $\alpha+\beta  = \pi \;(\mmod 2\pi)$, the correlated EPR state \eqref{eq:EPR} evolves to an anticorrelated EPR state
 \begin{equation}
\ket{\Phi^{\mathrm{EPR}}} = \iint d\rh_1 d\rh_2 \delta(\rh_1+\rh_2) \ket{\rh_1}_1\ket{\rh_2}_2.   
\label{eq:EPRanti}
\end{equation} 
In this case the detection of photon $2$ at $\rh$ projects photon $1$ onto the state $\ket{-\rh}$.   Given any propagation characterized by an FRFT $\mathcal{F}_{\alpha}$ on photon 1, one can find a transformation $\mathcal{F}_{\beta}$ on photon 2 such that a correlation or anticorrelation is recovered.  When  $\alpha+\beta  = \pi/2 \;(\mmod 2\pi)$, this state becomes  
 \begin{equation}
\ket{\Omega} = \int d\rh  \ket{\rh}_1\ket{\bq(\rh)}_2,   
\label{eq:EPR2}
\end{equation} 
where $\ket{\bq(\rh)}\propto \int \exp(i \bq(\rh)\cdot \rh)\ket{\rh}$ is the momentum eigenstate conjugate to $\ket{\rh}$.  State  \eqref{eq:EPR2} presents no  intensity correlation.  An equivalent result is found for $\alpha+\beta  = 3\pi/2 \;(\mmod 2\pi)$.  We note that the conditions for correlation, anti-correlation, and no-correlation depend on the sum of the FRFT angles of the down-converted fields, and not the individual angles $\alpha$ and $\beta$.    
\par
\begin{figure}
\centering
\includegraphics*[width=8cm]{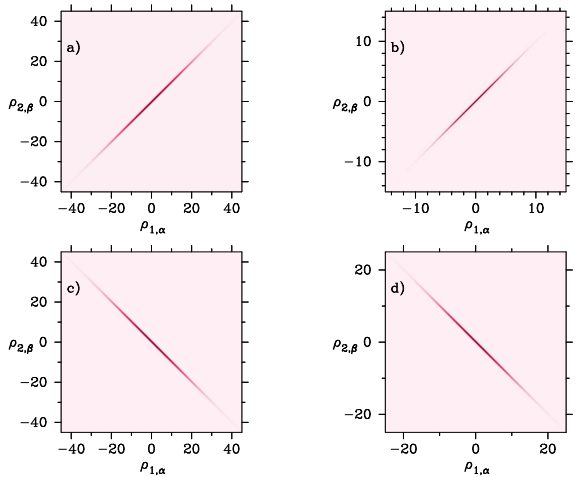}
\caption{\label{FIG:densityplot1} 
Density plot of the joint detection probability calculated for the initial gaussian
state in Eq.(\ref{EQ:TwopartWaveFuncPOS}) evolved with the following FRFT orders: a) $\alpha=3\pi/4$,
$\beta=5\pi/4$; b) $\alpha=\pi$, $\beta=\pi$; c) $\alpha=\pi/4$, $\beta=3\pi/4$ and  d) 
$\alpha=\pi/2$, $\beta=\pi/2$. A strong correlation is present when $\alpha+\beta=0 \;(\mmod 2\pi)$ [plots  a) and b)]
and a strong anti-correlation when $\alpha+\beta=\pi\;(\mmod 2\pi)$ [plots c) and d)]. }
\end{figure}
\begin{figure}
\centering
\includegraphics*[width=8cm]{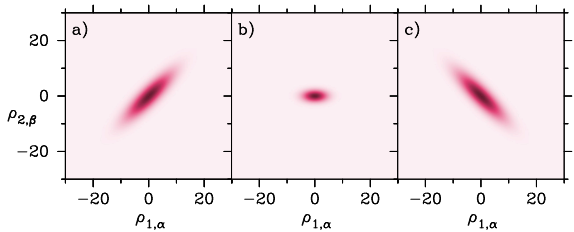}
\caption{ \label{FIG:densityplot2}
Density plot of the joint detection probability calculated for the initial gaussian
state in Eq.(\ref{EQ:TwopartWaveFuncPOS}) evolved with the following FRFT orders: a) $\alpha=\pi/4$,
$\beta=5\pi/4$; b) $\alpha=\pi/2$, $\beta=\pi$ and  c) $\alpha=3\pi/4$, $\beta=3\pi/4$.
In this case, with $\alpha+\beta=3\pi/2 \;(\mmod 2\pi)$, we have a strong decrease in  the correlations.}
\end{figure}

This simple picture drawn for the ideal EPR-state is followed approximately by the two-photon state in Eq.(\ref{EQ:TwopartWaveFuncPOS}).  For simplicity, let us use the fact that the two-photon wave function is factorable in $x$ and $y$ variables:  $\Psi(\rh_1,\rh_2)=\xi(\rho_{x_1},\rho_{x_2})\xi(\rho_{y_1},\rho_{y_2})$.  Then we can consider one spatial dimension $\rho$ for each down-converted field. 
Figures \ref{FIG:densityplot1} and \ref{FIG:densityplot2} show the initial state 
(\ref{EQ:TwopartWaveFuncPOS}) propagated under different FRFT's using $\sigma_+=4.076$ and $\sigma_-=0.067$.  FIG. \ref{FIG:densityplot1} shows two examples of strong correlations between photon 1 and 2 when the FRFT orders satisfy the condition $\alpha+\beta=0 \;(\mmod 2\pi)$ and two examples of strong anticorrelations  when  the condition is 
$\alpha+\beta=\pi\;(\mmod 2\pi)$.
For the condition  $\alpha+\beta=3\pi/2$, FIG. \ref{FIG:densityplot2} shows a significant decrease 
of intensity correlations, although in general they do not completely dissapear as is the case shown in Eq.(\ref{eq:EPR2}) for the ideal EPR state.
In fact, analytical calculation shows that,  in order to have no intensity 
correlation, {\it i.e.} $|\langle \rh_1,\rh_2|\Psi_{\alpha,\beta}\rangle|^2=f_{1}(\rh_1)f_{2}(\rh_2)$, the exact relation between $\alpha$ and $\beta$ is: 
 \begin{equation}
\tan\alpha\tan\beta =\sigma_{-}^2\sigma_{+}^2. 
\label{eq:no-intensity-correlation-condition}
\end{equation} 
Eq. \eqref{eq:no-intensity-correlation-condition} is satisfied by FRFT orders such that $\alpha+\beta=\pi/2 \;(\mmod 2\pi)$ or $\alpha+\beta=3\pi/2 \;(\mmod 2\pi)$
only when $\sigma_{-}=1/\sigma_{+}$. 
Nevertheless, the intensity correlations present in the state Eq. (\ref{EQ:TwopartWaveFuncPOS}) 
propagate in a fashion similar to idealized case of the EPR state.  
\par
In the laboratory, one has access to the joint detection probability, which in the case of a two-photon state corresponds to the fourth-order correlation function \cite{mandel95} 
\bea
P_{\alpha,\beta}(\rh_{1},\rh_{2})&=&\bra{\Psi_{\alpha,\beta}}\oper{a}^{\dagger}(\rh_1)\oper{a}^{\dagger}(\rh_2)\oper{a}(\rh_1)\oper{a}(\rh_2) \ket{\Psi_{\alpha,\beta}}\nonumber \\
&=&|\Psi_{\alpha,\beta}(\rh_1,\rh_2)|^2, 
\eea
and is proportional to the number of coincidence counts $C_{\alpha,\beta}(\rh_{1},\rh_{2})$.  The conditional probability can be obtained by the relation 
\begin{equation}
P_{\alpha,\beta}(\rh_{2}|\rh_{1}) = \frac{P_{\alpha,\beta}(\rh_{1},\rh_{2})}{P_{\beta}(\rh_{1})},
\label{eq:conditionalP}
\end{equation}
 where  $P_{\beta}(\rh_{1})$ is proportional to the number of single counts $C_{\beta}(\rh_{1})$. Thus, the conditional probability $P_{\alpha,\beta}(\rh_{2}|\rh_{1})$ is also  proportional to the number of two-photon coincidence counts $C_{\alpha,\beta}(\rh_{1},\rh_{2})$. 
%

\section{Experiment}
\label{sec:exp}
\begin{figure}[h]
\centering
\includegraphics*[width=8.5cm]{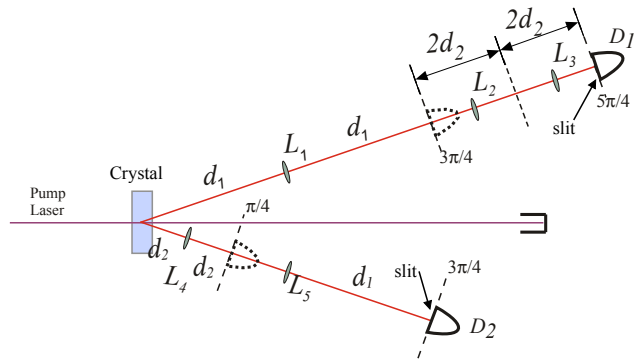}
\caption{\label{FIG:SETUP} (color online)  Experimental Setup. The distances of
the FRFT lens systems are $d_1=42.63$cm and $d_2=7.33$cm. All
lenses have focal length $f=25$cm. Moveable slits (not shown) are placed in front of each detector.}
\end{figure}
We investigated the propagation of EPR-like correlations experimentally by implementing several FRFT's on pairs of entangled photons and registering the coincidence counts while scanning one of the detectors.   
The experimental setup is shown in Fig. \ref{FIG:SETUP}. Degenerate twin photons with $\lambda=810$nm are generated by
pumping a 5mm long lithium iodate crystal (LiIO$_3$) with a $10$mW c.w.
diode laser centered at $\lambda_p=405$nm. The transverse waist of the beam at the laser output was measured to be $0.31\pm0.01$ mm.  To increase the spatial correlations, the beam width is expanded three times using two confocal lenses.  The down-converted photons are detected by APD
photodetectors equipped with $10$nm bandwidth interference filters centered at
$810$nm. Moveable horizontal  slits ($100\mu$m$\times3$mm) are placed directly in front of each  detector in order to scan the vertical
position. The FRFT's are performed on both down-converted fields using  the ``type I" symmetric lens system
configuration reported in \cite{lohmann93} and shown in FIG. \ref{FIG:FRFTs} b).  The dimensionless position and momentum
coordinates for this kind of system are
$\rh = \sqrt{{k}/{f^{\prime}}}{\rhb}$ and
$\bq = \sqrt{{f^{\prime}}/{k}}{\bqb}$, where $f^{\prime}=25/\sqrt{2}$ cm (see below) is the scaled focal length and $\rhb$, $\bqb$ are the dimensional variables.     
\par

\par
\begin{figure}
\centering
\includegraphics*[width=7cm]{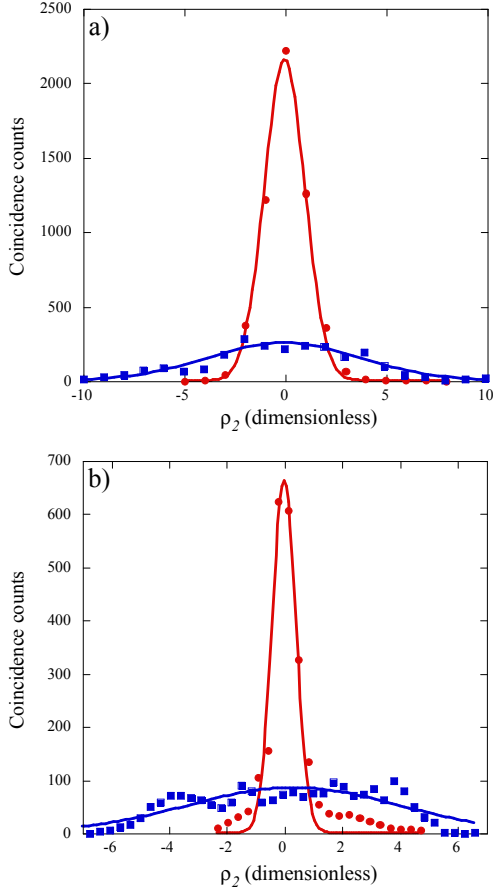}
\caption{\label{FIG:near-far} (Color online) Coincidence counts $C_{\alpha,\beta}(\rho_1,\rho_2)$ as a function of dimensionless $\rho_2$  for  a)  $\alpha=\pi, \beta=\pi$ (red circles) and $\alpha=\pi/2, \beta=\pi$ (blue squares) and b) $\alpha=\pi/2, \beta=\pi/2$ (red circles) and $\alpha=\pi, \beta=\pi/2$ (blue squares). In all cases detector 1 is fixed at $\rho_1=0$.}
\end{figure}
\par
Initially, correlation measurements for the near field ($\alpha=\beta=\pi$) and far-field ($\alpha=\beta=\pi/2$) correlations were obained by fixing one detector at $\rho=0$ and scanning the other \cite{howell04,almeida05}.    
These correlations correspond to the usual position and wave vector variables in the source plane.  The near field correlations were obtained by imaging the exit surface of the crystal on the plane of the detectors with $4f$ lens systems.  For the far-field measurements, the usual optical Fourier transform system was used.   A sample of the coincidence counts are shown in FIG.  \ref{FIG:near-far}, as a function of the dimensionless variable $\rho_2$.   The conditional variances are listed in table \ref{tab:variances}.  Using these results, we can evaluate the EPR inequality \eqref{EQ:EPRinequality}:
\begin{align}
      \Delta^2_{\pi,\pi}(\rho_{1}|\rho_{2})
      \Delta^2_{\pi,\pi}(q_{1}|q_{2})=0.20 \pm 0.01 <\frac{1}{4}\\
      \Delta^2_{\pi,\pi}(\rho_{2}|\rho_{1})
       \Delta^2_{\pi,\pi}(q_{2}|q_{1})=0.14 \pm 0.01<\frac{1}{4}, 
      \end{align}
which shows that the state displays non-local correlations.  Also shown in FIG.  \ref{FIG:near-far} are the results using different lens configurations, which give the weakly correlated intensity distributions.  For example, the $\pi-\pi/2$ distribution is more than 10 times larger than the $\pi-\pi$ and $\pi/2-\pi/2$ distributions.   
\begin{table}
  \begin{ruledtabular}
  \begin{tabular}{ccc}
  $\alpha,\beta$ & $\Delta^2_{\alpha,\beta}(\rho_{2}|\rho_{1})$ & 
  $\Delta^2_{\alpha,\beta}(\rho_{1}|\rho_{2})$ \\
  \hline
   $\alpha=\frac{\pi}{2}, \beta=\frac{\pi}{2}$ & 0.14 $\pm$ 0.02  & 0.17 $\pm$ 0.02  \\
         $\alpha={\pi}, \beta={\pi}$ & 0.98  $\pm$ 0.06 & 1.39 $\pm$ 0.06  \\
          $\alpha=\frac{\pi}{2}, \beta={\pi}$ & 12.3 $\pm$ 1.5&  $--$ \\
     $\alpha={\pi}, \beta=\frac{\pi}{2}$ & 13.3  $\pm$ 2.1 &  $--$ \\
  $\alpha=\frac{3\pi}{4}, \beta=\frac{5\pi}{4}$&  0.29 $\pm$ 0.01 & 0.28 $\pm$ 0.02 \\
   $\alpha=\frac{\pi}{4}, \beta=\frac{3\pi}{4}$ & 0.21 $\pm$ 0.01 & 0.31 $\pm$ 0.02 \\
    $\alpha=\frac{\pi}{4}, \beta=\frac{5\pi}{4}$ & 9.2   $\pm$ 0.7& 13.8  $\pm$ 1.2  \\
     $\alpha=\frac{3\pi}{4}, \beta=\frac{3\pi}{4}$ & 12.1 $\pm$  1.1  &17.7  $\pm$ 1.1  \\
       \end{tabular}
  \end{ruledtabular}
\caption{Conditional variances for all measurement results for different FRFT orders $\alpha$ and $\beta$.  Variances were obtained from gaussian curve fits.}\label{tab:variances}
\end{table}
\par
To evaluate the strength of these correlations under different FRFTs, a series of measurements were performed with various FRFT lens systems.  All lenses used in the experimental setup have the same focal length
$f=25$cm. We chose FRFT's with orders $\alpha= \{ \frac{3\pi}{4},\frac{5\pi}{4} \}$ and
$\beta=\{\frac{\pi}{4},\frac{3\pi}{4}\}$, where $\alpha$ and $\beta$ correspond to photons 1 and 2, respectively.  These FRFT orders sum to either $\pi$, $3\pi/2$, or $2 \pi$. This choice of angles is especially convenient, as it maintains $f^{\prime}=25/\sqrt{2}$ cm the same for all of the FRFT systems used.  This
is advantageous for several reasons: i) to respect the condition of
additivity of two consecutive FRFT's systems and ii) to use the same
scaling factor for signal and idler fields which is necessary in order to describe the FRFT mathematically as a rotation in phase space. The
scaling parameter for our system is
$\sqrt{{k}/{f^{\prime}}}=6.62$mm$^{-1}$.
\par
The various lenses used to implement these FRFTs are shown in FIG. \ref{FIG:SETUP}. Three additive
FRFT lens systems were used to perform the ${5\pi}/{4}$ order FRFT. Lens
$L_1$ is used to perform a ${3\pi}/{4}$ order FRFT of the
field from the exit face of the crystal to position
$2d_1$. Lenses $L_2$ and $L_3$ each perform a
${\pi}/{4}$ order FRFT, the first from $z=2d_1$ to
$z =2d_1+2d_2$ and the second from $z=2d_1+2d_2$ to
$z=2d_1+4d_2$. The field at the plane $z=2d_1+4d_2$ is the
FRFT of order ${5\pi}/{4}$ of the field at the exit
face of the crystal. Lens $L_5$  is used to perform a
$\frac{3\pi}{4}$ FRFT, and $L_4$ was used to perform a $\pi/4$ order FRFT.  By choosing different detector positions and combinations of lenses, we could implement several different FRFT's on each down-converted field.  
\par
\begin{figure}
\centering
\includegraphics*[width=7cm]{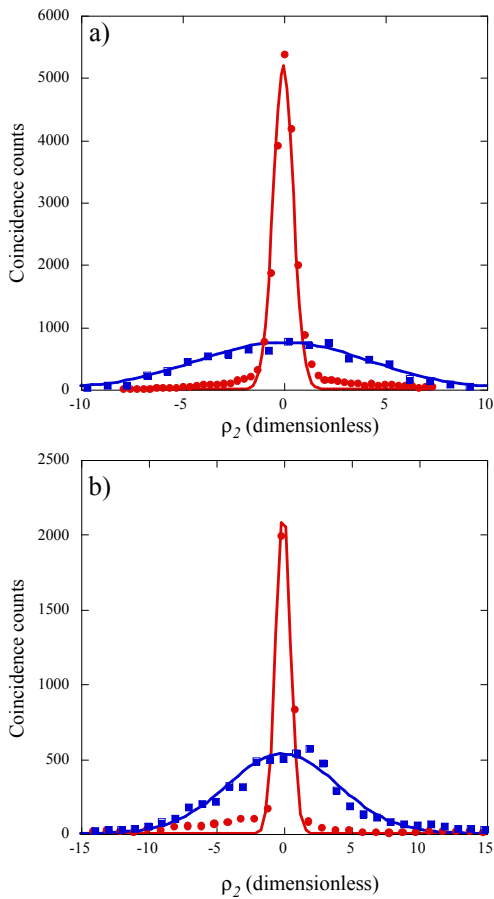}
\caption{\label{FIG:signal} (color online) Coincidence counts $C_{\alpha,\beta}(\rho_1,\rho_2)$ as a function of dimensionless $\rho_2$  for  a)  $\alpha=\pi/4, \beta=3\pi/4$ (red circles) and $\alpha=3\pi/4, \beta=3\pi/4$ (blue squares) and b) $\alpha=3\pi/4, \beta=5\pi/4$ (red circles) and $\alpha=\pi/4, \beta=5\pi/4$ (blue squares). In all cases detector 1 is fixed at $\rho_1=0$.}
\end{figure}

  
A sample of the experimental results are shown in FIG. \ref{FIG:signal}, which displays coincidence counts $C_{\alpha,\beta}(\rho_1,\rho_2)$ as a function of the dimensionless coordinate $\rho_2$.  In all of the plots the slit of detector 1 is fixed at the origin ($\rho_1$=0). These figures correspond to vertical cross-sections along the line $\rho_1=0$ of the theoretical density plots in FIG.'s \ref{FIG:densityplot1} and \ref{FIG:densityplot2}.  One can see that for the cases $\alpha+\beta=\pi, 2\pi$, a narrow coincidence distribution is observed, indicating either an intensity correlation or anti-correlation.  When  $\alpha+\beta=3 \pi/2$, the coincidence profile is much larger, indicating a much weaker correlation.  Using Eq. \eqref{eq:conditionalP}, the conditional variances $\Delta^2_{\alpha,\beta}(\rho_{2}|\rho_{1})$ were determined through gaussian curve fits of the coincidence distributions. Similar measurements and analysis were conducted by scanning $\rho_1$ and fixing detector 2 at $\rho_2=0$. The dimensionless variances for all results obtained are presented in table \ref{tab:variances}.  We note that the variance for the weakly correlated distributions are about 10-50 times larger than the correlated and anti-correlated distributions.      

\par
The coincidence distribution $C_{\frac{\pi}{4},\frac{3\pi}{4}}(\rho_{1}|\rho_{2})$ in fact corresponds to the the transverse wave-vector
distribution $C_{\frac{3\pi}{4},\frac{5\pi}{4}}(q_{1}|q_{2})$, since the $\alpha=\pi/4$ ($\beta=3\pi/4$) FRFT differs from the $\alpha=3\pi/4$ ($\beta=5\pi/4$) FRFT by a Fourier transform.  
Thus, with the experimental results shown in Fig. \ref{FIG:signal}, we can calculate the EPR inequality \eqref{EQ:EPRinequality}:  
    \begin{equation}
      \Delta^2_{\frac{3\pi}{4},\frac{5\pi}{4}}(\rho_{1}|\rho_{2})
      \Delta^2_{\frac{3\pi}{4},\frac{5\pi}{4}}(q_{1}|q_{2})=0.0391 <\frac{1}{4}, 
        \label{EQ:RESULTS}
        \end{equation}
  indicating EPR nonlocality.  Similiarly, the conditional variances $\Delta^2_{\alpha,\beta}(\rho_{1}|\rho_{2})$ give
      \begin{equation}
      \Delta^2_{\frac{3\pi}{4},\frac{5\pi}{4}}(\rho_{2}|\rho_{1})
       \Delta^2_{\frac{3\pi}{4},\frac{5\pi}{4}}(q_{2}|q_{1})=0.0352<\frac{1}{4}. 
       \label{EQ:RESULTS2}
    \end{equation}
    It is clear that the EPR intensity correlation is lost when $\alpha+\beta = 3 \pi/2$, since 
    \begin{align}
      \Delta^2_{\frac{3\pi}{4},\frac{3\pi}{4}}(\rho_{1}|\rho_{2})
      \Delta^2_{\frac{3\pi}{4},\frac{3\pi}{4}}(q_{1}|q_{2})=244 \pm 26  > \frac{1}{4}\\
      \Delta^2_{\pi,\pi}(\rho_{2}|\rho_{1})
       \Delta^2_{\pi,\pi}(q_{2}|q_{1})= 111 \pm 13 > \frac{1}{4}.  
      \end{align}

 \begin{figure}
\centering
\includegraphics*[width=7cm]{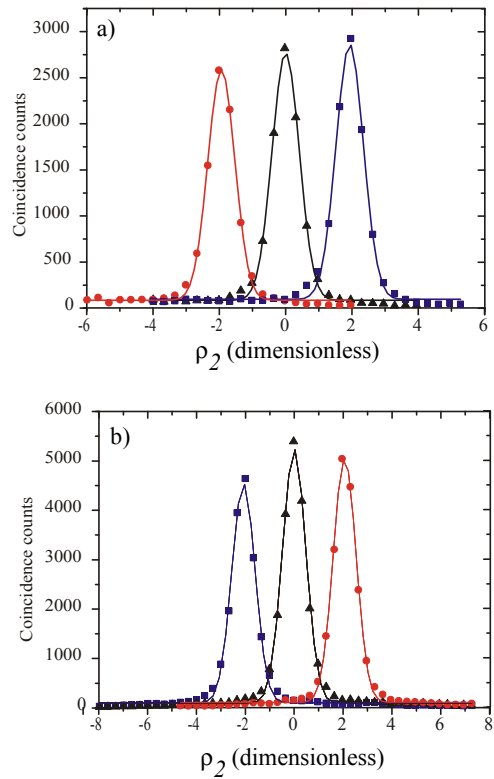}
\caption{\label{FIG:correlation} Coincidence counts $C_{\alpha,\beta}(\rho_1,\rho_2)$ as a function of dimensionless $\rho_2$ for a) $\alpha = 3\pi/4$ and $\beta=5\pi/4$ and  and b)  $\alpha = \pi/4$ and $\beta=3\pi/4$.  In both figures, the black triangles correspond to $\rho_1=0$, the red circles to $\rho_1=-1.99\pm0.03$ and the blue squares to $\rho_1=1.99\pm0.03$. Figure a) thus shows a correlation, while b) shows  an anti-correlation of the detection positions.}
\end{figure}
\par
The results summarized in table \ref{tab:variances} show the strength, but do not indicate the type of correlation.  To investigate the type of spatial correlation
in intermediate FRFT planes, we first used lens configurations with FRFT orders $\alpha=\frac{5\pi}{4}$ and
$\beta=\frac{3\pi}{4}$,  satisfying
$\alpha+\beta=2\pi$.  Experimental results are shown in FIG. \ref{FIG:correlation} a).  
Initially, the slit in front of
detector 1 was placed at the origin ($\rho_1=0$) and the slit in front of
detector 2 was scanned vertically. The measured coincidence counts are plotted in black
triangles in figure \ref{FIG:correlation} a) and the maximum
of the gaussian fit is at $\rho_1=0\pm0.03$. We then displaced one of the slits by $300\pm5\mu$m,
which corresponds to a dimensionless displacement of $\rho_1 = 1.99\pm0.03$.
Coincidence counts were again measured while the slit of
detector 2 was scanned.   Coincidence counts are in blue squares in FIG.
\ref{FIG:correlation} a) and the maximum of the gaussian fit is at
$\rho_2=1.93\pm0.03$.  Slit 1 was then moved $-300\pm5\mu$m,  ($\rho_1 = -1.99\pm0.03$), and slit 2 was scanned. The maximum of the coincidence counts occurred at $\rho_2=-1.94\pm0.03$. We thus observe
a strong correlation between the transverse coordinates for this
configuration satisfying $\alpha+ \beta=2\pi$. The same procedure was performed for the lens
configuration $\{\alpha=\frac{3\pi}{4}, \beta=\frac{\pi}{4}\}$, which satisfies the anti-correlation condition
$\alpha+\beta=\pi$.  The results are shown in FIG. \ref{FIG:correlation} b).   We observe similar displacement of the coincidence peaks, however in this case 
the maxima of the gaussian fits are anti-correlated with the position of the slit of the
other detector.    
\par
 \section{Conclusion}
\label{sec:conclusion}
We have used the fractional Fourier transform to study the propagation of the transverse intensity correlations of the two-photon state produced from parametric down-conversion.  The transforms were implemented with simple lens systems. Our theoretical and experimental results show that the propagation of the transverse correlations of highly-correlated two-photon states depends upon the the sum of the transform orders of the down converted fields.  For $\alpha + \beta = 0 \;(\mmod 2\pi)$, the original intensity correlation at the source is recovered, while for $\alpha + \beta = \pi\;(\mmod 2\pi)$, an intensity anti-correlation is observed.  For $\alpha + \beta= \pi/2 \;(\mmod 2\pi)$ or  $\alpha + \beta= 3\pi/2 \;(\mmod 2\pi)$, almost no correlation is present.  Analytical results were obtained for the propagation of the ideal EPR state, and numerical calculations along with our experimental results show that the down-converted photons display a similar behavior.  The EPR correlation present in the two-photon state was confirmed for several different orders of the fractional Fourier transforms through violation of an inequality.  These results apply to spatially correlated photons obtained from any source, as well as correlations present in other physical systems, and should be useful for engineering spatial correlations, as well as fundamental studies of quantum nonlocality and entanglement.              

\begin{acknowledgements}
We would like to thank A. Salles for fruitful discussions.
Financial support was provided by Brazilian agencies CNPq, PRONEX,
CAPES, FAPERJ, FUJB and the Milenium Institute for Quantum
Information.
\end{acknowledgements}

\bibliographystyle{apsrev}


\end{document}